\documentstyle[preprint,aps]{revtex}
\begin{document}
\draft
\preprint{}
\title{Does Magnetic Charge Imply a Massive Photon ?}
\author{D. Singleton}
\address{Department of Physics, Virginia Commonwealth University,
Richmond, VA 23284-2000}
\date{\today}
\maketitle
\begin{abstract}
In Abelian theories of monopoles the magnetic charge is
required to be enormous. Using the electric-magnetic duality of
electromagnetism it is argued that the existence of such a
large, non-perturbative magnetic
coupling should lead to a phase transition
where magnetic charge is permanently confined and the photon
becomes massive. The apparent masslessness of the photon could
then be used as an argument against the existence of such a large,
non-perturbative magnetic charge. Finally it is shown that even in the
presence of this dynamical mass generation the Cabbibo-Ferrari
\cite{ferr} formulation of magnetic charge gives a consistent
theory.
\end{abstract}
\newpage
\narrowtext

\section{Strong Coupling Phase Transition}

Normally the gauge bosons of a theory are said to be massless due
to the requirement of gauge invariance. If the Lagrangian of a theory
has a mass term for the gauge bosons ({\it i.e.} a term like ${1 \over 2}
m^2 A_{\mu} A^{\mu}$) then the Lagrangian is no longer invariant under
the gauge transformation of the gauge field ({\it i.e.} $A_{\mu}
\rightarrow A_{\mu} - {1 \over e}
\partial_{\mu} \Lambda (x)$, where $\Lambda (x)$
is an arbitrary function ). One caveat to this prohibition is the Higgs
mechanism \cite{higgs} which allows the gauge boson to have a mass
while still remaining consistent with gauge invariance, by coupling
the gauge boson to a scalar field which develops a vacuum expectation
value. A less often stated caveat is that the coupling of the gauge
boson to particles of the theory needs to be small enough \cite{huang}
so that the gauge boson does not become massive through some
non-perturbative mechanism ({\it e.g.} techni-color models for
mass generation in the standard model). It is diffiuclt to give
a definite value for how small the coupling constant should be in
order to insure the masslessness of the gauge boson, but requiring that
it be small enough so that perturbation theory is valid seems a good
rule of thumb. Wilson has argued \cite{wilson} that in a U(1) gauge
theory there should be some critical coupling, $e_c$, below which the
U(1) gauge boson is massless and above which it acquires a mass. Wilson's
conjecture does not determine whether this phase transition from massless
gauge boson to massive gauge boson is a first or second order
transition, nor does it give the value of the critical coupling at which
this transition should occur. This conjectured mechanism, which
dynamically generates a mass for the U(1) gauge boson, is similiar to
an effect which was found to occur in QED in $1+1$ dimensions.
Schwinger \cite{schwing} rigorously showed that in $1+1$ dimensions
the photon would acquire a mass proportional to $e^2$, the square of
the coupling. Thus in $1+1$ dimensional QED $e_c = 0$, and
the photon always becomes massive. Schwinger also conjectured
that the same effect could occur in $3+1$ QED for some unspecified,
large coupling. Guth \cite{guth} has shown that a U(1) gauge theory
will indeed undergo a phase transition as
conjectured by Wilson and Schwinger, but no
theoretical value for the critical coupling constant was given.
Thus for $3+1$ dimensional QED it may be an ``accident''
of the gauge coupling, $e$, being small that results
in the physical photon being massless within very stringent
limits (the upper bound on the photon mass is
$3.0 \times 10^{-27} eV = 5.3 \times 10 ^{-63} kg
> m_{\gamma}$ \cite{pdhb}). The amazing success of perturbation theory
for the electromagnetic interactions of the electron also indicates
that the physical electromagnetic coupling is below this unknown
critical value. QCD in contrast is thought to exist in the confining
phase with a fine structure constant, $\alpha _s = {g_s ^2 \over 4 \pi}$
on the ${\cal O} (1)$.

In Dirac's theory of magnetic charge one allows the vector
potential {\bf A} to develop a singularity  that runs from
the location of the magnetic charge to spatial infinity, so
that $\nabla \cdot {\bf B} = \rho _m$ is consistent
with the ${\bf B} = \nabla \times {\bf A}$
\cite{dirac}. Dirac also showed that in order for the
wavefunction of an electrically charged particle in the
presence of this string singularity to be single valued, the
following quantization condition had to hold
\begin{equation}
\label{dirac}
{e g \over 4 \pi} = {n \over 2}
\end{equation}
Where $n$ is an integer, $g$ is the magnitude of the magnetic
charge and $e$ is the magnitude of the electric charge (which
we will take to be the charge of the electron). There are other
ways of formulating a theory of magnetic charge without
having to take recourse to a singular vector potential
(the fiber bundle approach of Wu and Yang \cite{wu} or the
two-potential approach of Cabbibo and Ferrari \cite{ferr}).
In all these various theories, however, one eventually
ends up with a similiar quantization condition. The best
model independent argument for this is due to Saha
\cite{saha}. If one considers a particle with electric charge
$e$ in the presence of a particle with magnetic charge
$g$, then due to the ${\bf E} \times {\bf B}$ term in the
energy-momentum tensor this system carries a
field angular momentum
of magnitude $eg / 4 \pi$. Since angular momentum is quantized
in integer multiples of $\hbar /2$ we again arrive at
condition Eq. (\ref{dirac}), where we have set $\hbar = 1$.

If $e$ in Eq. (\ref{dirac}) is taken as the physical charge
of the electron it is found that the magnitude of the the
magnetic charge is enormous. The strength of the electric coupling
strength between two electric charges is ${e^2 \over 4 \pi}
\approx {1 \over 137}$, while the strength of the minimum
magnetic coupling ({\it i.e.} $n = 1$ in Eq. (\ref{dirac}))
between two monopoles is ${g^2 \over 4 \pi} \approx {137 \over 4}$.
The interaction strength between two monopoles is roughly
$5 \times 10 ^3$ times stronger than between two electric
charges. The size of the magnetic coupling puts it well out of
the range of perturbation theory, and opens up the logical possibility
that unusual non-perturbative effects could occur in the presence of
such a non-perturbative magnetic charge. In Wilson and Guth's argument
for a phase transition in a U(1) gauge theory with a large coupling,
the U(1) gauge charge is usually thought of as electric charge.
If the U(1) gauge charge is taken to be electric charge then there
is a definite difference, in the standard formulation of the theory,
in the way the gauge boson couples to electric charge as
compared to how it couples to magnetic charge. The electric charge is
minimally coupled to the vector potential,  $A_{\mu}$, while the
magnetic charge has no simple coupling to $A_{\mu}$. Physically,
however, the gauge boson should couple to both charges in
a symmetric way, especially when one looks at the theory
in terms of how these charges interact with the {\bf E} and
{\bf B} fields. This physically intuitive idea takes the mathematical
form of a dual symmetry between electric and magnetic charges,
which indicates that the two types of charges are indeed
interchangeable. This dual symmetry is  \cite{jackson}
\begin{eqnarray}
\label{dual}
J_e ^{\mu} \rightarrow J_e ^{\mu} cos \theta + J_m ^{\mu}
sin \theta \nonumber \\
J_m ^{\mu} \rightarrow -J_e ^{\mu} sin \theta + J_m ^{\mu}
cos \theta
\end{eqnarray}
where $J_e ^{\mu} = (\rho _e , {\bf J} _e)$ and $J_m ^{\mu} =
(\rho _m , {\bf J} _m)$ are the electric and magnetic four
current densities respectively. This dual symmetry between
electric and magnetic charges and currents shows that it is
a matter of convention as to what is called electric charge
and what is called magnetic charge.
In fact Baker {\it et. al.} \cite{ball} have shown
that electromagnetism can be reformulated with magnetic
charge as the gauge charge, which is then minimally coupled to
the U(1) gauge boson, while electric charge  is attached to
Dirac type strings. Combining this dual symmetry with the
strong coupling phase transition to a confining
phase with a massive gauge boson, it can be argued that a large,
non-perturbative magnetic charge would make the photon massive.
The dual symmetry is important since it
indicates that it should not make a difference
whether the large, non-perturbative charge is electric or magnetic.
Since the photon is apparently massless to some stringent upper limit
\cite{pdhb} this implies that Abelian
magnetic charge is absent from the
physical world. As we shall see the Cabbibo-Ferrari formulation
of magnetic charge could still give a consistent theory even
in the presence of this dynamical mass generation. Even
though there is no theoretical prediction as to the critical value
of the coupling at which this phase transition should occur, the
value at which QCD apparently undergoes this phase transition,
while not exactly determined, is certainly thought to be much
less than $137/4$. Finally numerical work on compact lattice
U(1) gauge theory points to a critical coupling of the order
unity \cite{tousi}. Assuming that as the limit of the lattice
spacing is taken to zero that the lattice theory goes over
smoothly into the continuum theory one again finds an indication
that the required value of the magnetic coupling is in the
confinement regime where the gauge boson is massive.

\section{The Technicolor Analogy}

In this section we will give an argument, based on
an analogy to the technicolor idea, that also points to the
possibility that in the presence of magnetic charge the photon
would develop a dynamical mass. The basic idea behind technicolor
theories is to introduce a new set of fermions ({\it i.e.} techni-
fermions) which couple to a new strong, non-Abelian gauge force
called technicolor. The techni-fermions form a condensate,
$\langle {\bar F} F \rangle \ne 0$, which gives the theory a
vacuum expectation value. The elementary Higgs scalar is
replaced by the composite scalar, ${\bar F} F$, which must have the
correct quantum numbers in order to mix with the gauge
boson that is to become massive. In the present case
instead of the composite scalar being composed of techni-fermions
it is composed of a monopole-antimonopole pair. Denoting the
monopole-antimonopole condensate by $\Pi _m$ we can, in analogy
with technicolor introduce an effective coupling between the
photon and this composite scalar particle
\begin{equation}
\label{interact}
{\cal L} _{\gamma - m} =  {f_m \over 2} (g A^{\mu})
(\partial _{\mu} \Pi _m)
\end{equation}
Where $f_m$ is a constant, which
is the equivalent of the pion decay constant of QCD.
This interaction term in the Lagrangian mixes the photon with
the composite $\Pi _m$ with a Feynman rule vertex  of
$ - {i g f_m \over 2} q_{\mu}$, were $q_{\mu}$ is the momentum
of the photon. Taking an infinite sum of $\Pi _m$ 's mixing in
with the photon changes the photon's propagator from
\begin{equation}
\label{prop1}
D _{\mu \nu} ^{\gamma} = {-i(g^{\mu \nu} -q_{\mu} q_{\mu} / q^2) \over
q^2}
\end{equation}
to
\begin{equation}
\label{prop2}
D_{\mu \nu} ^{\gamma} = {-i (g^{\mu \nu} -q_{\mu} q_{\mu} / q^2) \over
q^2 - g^2 f_m ^2 / 4}
\end{equation}
The pole in the second propagator indicates that that photon now
has a mass of $m_{\gamma} = g f_m /2$. This mass is arbitrary since
the ``magnetic'' pion decay constant, $f_m$, is unspecified. Both
the argument based on Wilson and Guth's idea of a phase transition
for a strongly coupled theory, and this more heuristic techni-color
inspired argument point to the photon developing a mass in
the presence of a large magnetic charge. Both arguments
have a degree of ambiguity. In the first case the critical value
at which the phase transition occurs is not determined theoretically,
although QCD apparently undergoes such a phase transition at a value
of the coupling which is much less than the required
magnitude of magnetic coupling. In the
second case the mass given to the photon is arbitrary since it
depends on the unknown ``magnetic'' pion decay constant, $f_m$.
In either case one could still make
the argument that the mass given
to the photon by the non-perturbative magnetic charge is smaller
than the experimental upper limit on the photon mass. Given
the stringent upper bound on the photon mass this argument is
unnatural. The more likely statement is that the apparent
masslessness of the photon implies the absence of magnetic charge.

\section{Discussion and Conclusions}

Using two different approaches we have argued that the required
large, non-perturbative value of magnetic charge is inconsistent
with the apparent masslessness of the photon. Or put in reverse :
the apparent masslessness of the photon implies the absence of
magnetic charge with the large, non-perturbative coupling which is
required in monopole theories.
This statement is too broad. The Cabbibo-Ferrari
formulation \cite{ferr} of magnetic charge could still remain
consistent with this dynamical  mass generation for the photon
{\it if} one intreprets the second potential as a second gauge
boson. In the Cabbibo-Ferrari approach a second pseudo four-vector
potential $C_{\mu} = (\phi _m , {\bf C})$ is introduced in addition
to the usual four-vector potential $A_{\mu} = (\phi _e , {\bf A})$.
Then in terms of these two potentials the normal definitions of the
{\bf E} and {\bf B} field get expanded to
\begin{equation}
\label{eb}
E_i = F^{0i} - {\cal G}^{0i} \; \; \; \; \; B_i = G^{0i} +
{\cal F} ^{0i}
\end{equation}
where the field strength tensors are
\begin{equation}
\label{fst}
F_{\mu \nu} = \partial _{\mu} A_{\nu} - \partial _{\nu} A_{\mu}
\; \; \; \; \;
G_{\mu \nu} = \partial _{\mu} C_{\nu} - \partial _{\nu} C_{\mu}
\end{equation}
and their duals
\begin{equation}
\label{dfst}
{\cal F} _{\mu \nu} = {1 \over 2} \epsilon _{\mu \nu \rho \sigma}
F^{\rho \sigma} \; \; \; \; \; \;
{\cal G} _{\mu \nu} = {1 \over 2} \epsilon _{\mu \nu \rho \sigma}
G^{\rho \sigma}
\end{equation}
Even though there are two potentials in this approach one normally
imposes conditions on these two potentials so that in the end
there are only enough degrees of freedom left to account for one
photon \cite{zwang}. In the Cabbibo-Ferrari theory one
also ends up with an enormous, non-perturbative value for the
magnetic coupling due to Saha's angular momentum quantization
argument. Thus, in the one photon version of the
Cabbibo-Ferrari formulation, the apparent observed masslessness
of the photon again implies the absence of
magnetic charge. If, however, the pseudo four-vector potential
is taken to be a second, parity odd photon then a consistent
theory can be given even in the presence of a large, non-perturbative
magnetic coupling. One can arrange for the dynamical symmetry
breaking to give a mass to the pseudo photon, $C_{\mu}$, while
the second photon, $A_{\mu}$, remains massless. This is in direct
analogy with what happens in the $SU_L (2) \times U(1)$ standard
model, where the $Z$ boson becomes massive while the photon
remains massless. This happens whether the symmetry breaking is
spontaneous or dynamical. Thus taking, $C_{\mu}$, as a real
gauge boson not only allows one to have a non-perturbative magnetic
coupling, but also naturally explains the absence of this second
pseudo photon from the particle spectrum that has so far been probed.
Most work on the Cabbibo-Ferrari theory of magnetic charge takes
the view of Ref. \cite{zwang} that there is only one photon. However
there are a few works which do regard the potential, $C_{\mu}$, as
being a second, physical photon \cite{salam}.

Wilson and Guth have argued that in a U(1) gauge theory there
should be a critical value of the coupling such that the theory
undergoes a phase transition to a confining theory  where the U(1)
gauge boson becomes massive. Combining this idea with the required
large, non-perturbative magnetic charge which occurs in all
monopole theories, and the electric-magnetic duality
(which implies that it should not matter whether the
non-perturbative coupling is electric or magnetic) we contend
that the photon acquires a dynamical mass in the presence of
magnetic charge. From an experimental point of view one can
point to the SU(3) theory of the strong interaction,
which is thought to exist in the confining phase with a
coupling constant that is considerably less then the coupling
constant a magnetic monopole is required to have.
The apparent experimental masslessness of the
photon then implies the absence of Abelian magnetic monopoles
of the Dirac or Wu-Yang type. A consistent monopole theory is
still possible if one works with the Cabbibo-Ferrari
theory {\it and} takes the somewhat unorthodox view that
the second pseudo four-vector potential corresponds to a physical
gauge boson.

The arguments given here should be taken strictly as applying only
to Abelian monopoles. Objects like the 't Hooft-Polyakov monopole
\cite{poly}, while also having an enormous magnetic charge, are of
a somewhat different character than the Dirac or Wu-Yang monopoles.
These magnetically charged objects come from an embedding
of a U(1) symmetry within a larger non-Abelian gauge group.
Additionally the magnetic charge of the theory is connected
with the unusual topological structure of the Higgs field. Both of
these facts make it difficult to formulate an electric-magnetic dual
symmetry for the 't Hooft-Polyakov theory. Since this dual symmetry
was crucial to our argument we can not use the
arguments presented here to place any restrictions
on the existence of 't Hooft-Polyakov magnetic charges.

\section{Acknowledgements} The author wishes to thank Siegfried
Roscher and Leonard O' Neill for help and encouragement during
the completion of this work. Additionally the author acknowledges
useful conversations with Atsushi Yoshida and Mike Timmins.

\end{document}